# Transport in a Single Self-Doped Nanocrystal

Hongyue Wang,[1] Emmanuel Lhuillier,[2] Qian Yu,[1] Alexandre Zimmers,[1] Benoit Dubertret,[1] Christian Ulysse,[3] and Hervé Aubin[*1]

[1]. ESPCI-ParisTech, PSL Research University, UPMC Univ. Paris 06, 10 rue Vauquelin
[1] LPEM, CNRS, F-75231 Paris Cedex 5, France
[2] Sorbonne Universités, UPMC Université Paris 06, CNRS-UMR 7588, Institut des NanoSciences de Paris, F-75005 Paris, France
[3] Centre de Nanosciences et de Nanotechnologies, CNRS, Univ. Paris-Sud, Université Paris-Saclay, C2N – Marcoussis, 91460 Marcoussis, France

**Abstract**: Addressing the optical properties of a single nanoparticle in the infrared is particularly challenging, thus alternative methods for characterizing the conductance spectrum of nanoparticles in this spectral range need to be developed. Here we describe an efficient method of fabricating single nanoparticle tunnel junctions on a chip circuit. We apply this method to narrow band gap nanoparticles of HgSe, which band structure combine the inverted character of the bulk semimetal with quantum confinement and self-doping. Upon tuning the gate bias, measurement reveals the presence of two energy gaps in the spectrum. The wider gap results from the interband gap, while the narrower gap results from intraband transitions. The observation of the latter near zero gate voltage confirms the doped character of the nanoparticle at the single particle level, which is in full agreement with the ensemble optical and transport measurements. Finally we probe the phototransport within a single quantum dot and demonstrate a large photogain mechanism resulting from photogating.

Electronic transport measurements of single nanoparticles and single molecules has been mostly employed for tunnel spectroscopy studies of fundamental phenomena such as the Kondo effect in single molecules,[1] the superconducting parity effect in small nanoparticles,[2] the electronic levels distribution in gold nanoparticles,[3] the electron-phonon coupling in PbS QDots[4] or non-equilibrium transport in magnetite nanocrystals.[5] Tunnel spectroscopy offers a complementary approach to optical measurements to probe the electronic structure of confined semiconductor nanostructures. Among them, 0D structures such as Quantum dots (QDot) have attracted a significant interest.[6] However, obtaining reliable and stable tunnel junctions with single QDots remains a difficult challenge.[7]

So far, most of the efforts regarding tunnel spectroscopy of colloidal QDots have been focused on wide band gap materials,[8-10] where the introduction of quantum confinement marginally affects the electronic structure. On the other hand, in narrow band gap semiconductors or semimetals, the quantum confinement leads to a drastic modification of the electronic spectrum. This is in particular the case for mercury chalcogenides, HgTe and HgSe, which have large Bohr radius, $a_0$=40 nm[11] and $a_0$= 33 nm (see SI) respectively. They are semimetals with an inverted band structure as bulk material. In this case, quantum confinement not only leads to a discrete electronic spectrum but is also expected to lead to a reordering of the bands.[12] Thus, an experimental determination of the electronic structure of mercury chalcogenide QDots is of interest.

Since these materials have a limited band gap (hundreds of meV) they present optical features in the mid-infrared (IR), which generated a strong interest for the fabrication of infra-red detectors.[13-15] For this application, the broadening of the spectrum is an essential question, which is generally addressed using single nanoparticle optical measurements. However, single nanoparticle optical spectroscopy become almost impossible in the IR and alternative single nanoparticle methods need to be developed to understand what are the current limiting factor regarding the electronic states linewidth (homogeneous vs inhomogeneous origin).

In this paper, we demonstrate that tunnel spectroscopy through a single HgSe QDot can reliably be measured. The choice of HgSe has been further motivated by its self-doped character[16,17] which leads to the presence of a few carriers within the dot. This doping actually results from the combination of the large work function of HgSe[18] (>6eV) and the narrow band gap nature of HgSe nanocrystals, which bring the conduction band below the $O_2/H_2O$ redox couple. As a result, water becomes a reducing agent for the HgSe QDot and its stable form is negatively charged.[19] Moreover since the method is based on electronic transport, it can also elucidate whether transport occurs through the QDot quantum states or rather by surface states.[20] To do single nanoparticle spectroscopy as function of a gate voltage, we developed an optimized electrospray method for the deposition of nanoparticle within pre-formed nanogap electrodes and conduct on chip tunnel spectroscopy. Compared to conventional scanning tunnel spectroscopy, this method offers two main advantages which are (*i*) a higher stability, and (*ii*) the possibility to add a gate for *in situ* control of the carrier density. Finally we probe the photoresponse of a single nanoparticle. In addition to the use of this material for photodetection, they may find application for the electrical detection of magnetic resonance.[21,22]

## Discussion

HgSe has a peculiar inverted band structure, see Figure 1a. At the Brillouin zone center, the Hg *s* level that forms a state of $\Gamma_6$ symmetry is pulled down below the anionic *p*-like $\Gamma_8$ level due to the large effective positive charge of the Hg core. Because the number of valence electrons is sufficient to occupy only two of the four levels of $\Gamma_8$ character, the unoccupied $\Gamma_8$ levels become part of the conduction band, which consequently becomes degenerate with the valence band maximum at Γ point, creating a zero energy gap.[23-26]

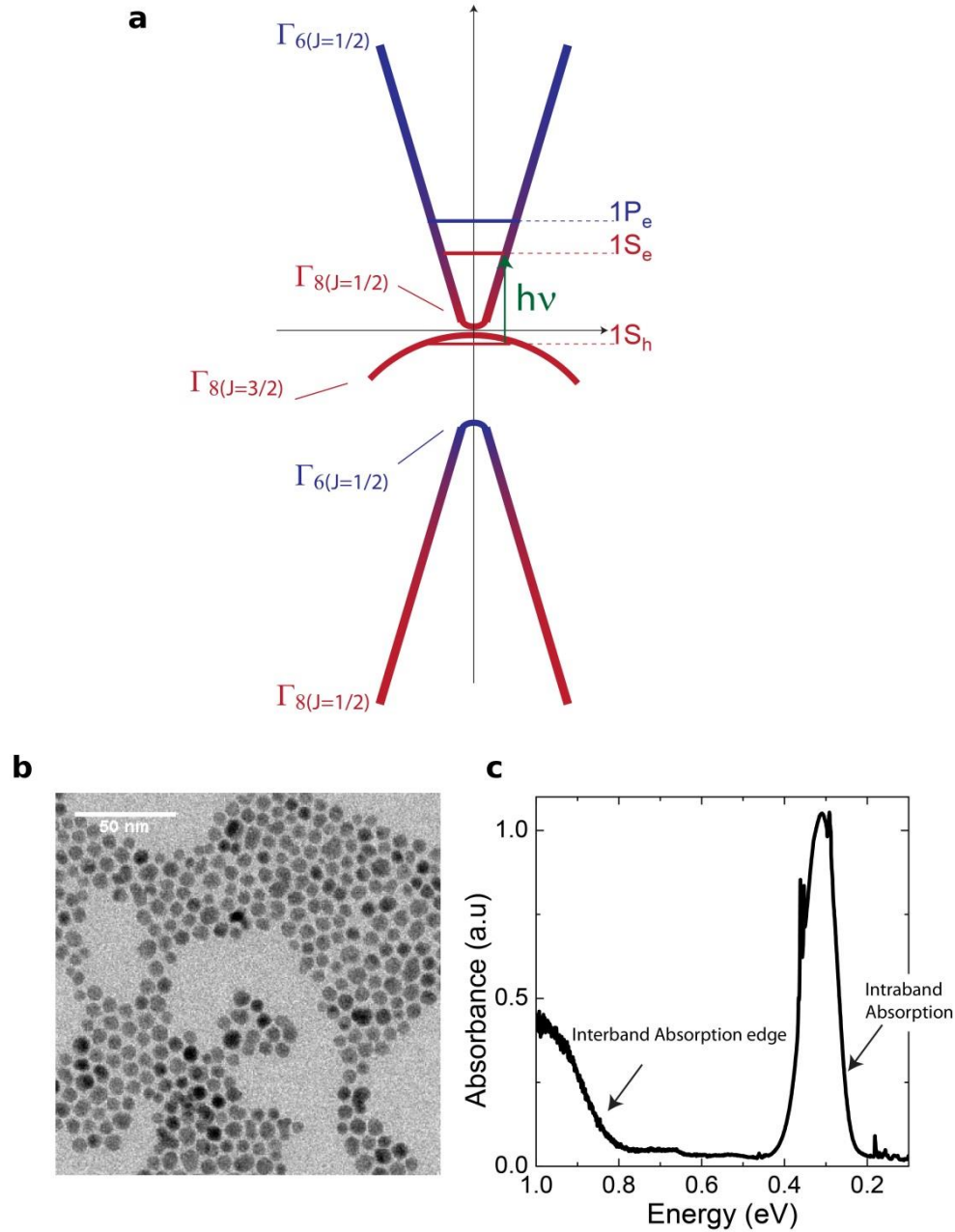

Figure 1: a) Sketch of the band diagram expected for HgSe QDots. At the zone center, the character $\Gamma_6$ (blue) and $\Gamma_8$ (red) of the bands is reversed with respect to the normal band ordering. Moving away from the zone center, the bands recover their normal ordering. b) TEM image of HgSe QDots. c) Optical absorption spectrum of HgSe QDots. The intraband gap is 2000 cm$^{-1}$ (0.25 eV), the interband gap is 6850 cm$^{-1}$ (0.85 eV).

Upon quantum confinement, the discrete electron (hole) levels move up (down) in the band structure, as sketched in Figure 1a, leading to finite interband and intraband gaps. The nanocrystals, see TEM image in Figure 1b, are synthesized according to the procedure given in Reference 17 and described in the method. Their diameter is about 10±2 nm. The optical absorption spectrum, shown in Figure 1c, displays an absorption edge at ≈ 0.25 eV due to intraband transitions from the 1Se to 1Pe levels. Furthermore, an absorption edge is also observed at ≈ 0.85 eV due to interband transitions from $1S_h$ to $1S_e$ levels.

To fabricate a single QDot device, we employed a method recently developed in the group[27] whose working principle is based on the projection of the QDots into a high vacuum chamber ($10^{-6}$ mbar), see the sketch in Fig. 2a. In this chamber, a chip circuit containing 32 nanogaps, also shown in Fig. S1, is located on the path of the QDots beam. One nanogap, Fig. 2c, is constituted by two electrodes separated of a distance ≈ 10 nm. In our past works,[Erreur ! Signet non défini.,27] the nanoparticles were projected with a pulsed valve. Since then, we improved the setup by replacing the pulse valve with an electrospray (model UHV4, ultra-high vacuum compatible, from MolecularSpray Ltd). This change leads to a higher success ratio and a cleaner on chip deposition with less organic residual.

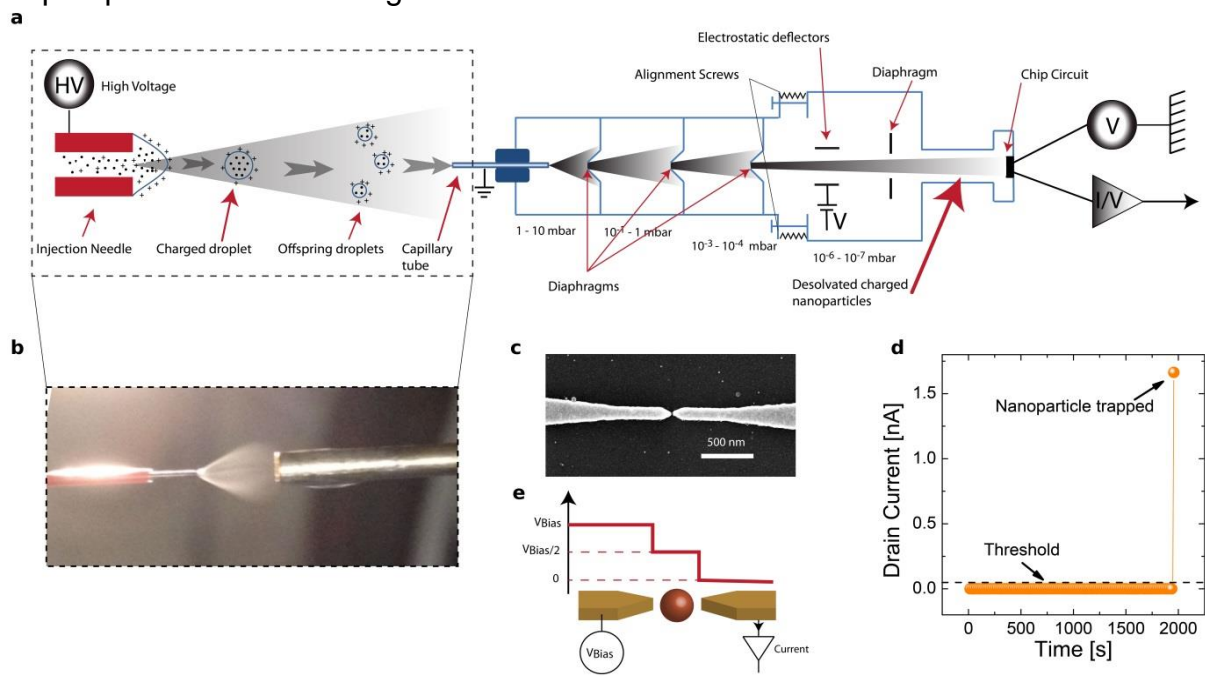

Figure 2 a) Sketch of the electrospray setup employed to fabricate the single QDot junctions. b) Picture of the spray induced by the high voltage (≈3000 V) between the injector needle and the capillary tube. After being injected into the capillary, the solvated beam of QDots goes through several chambers of decreasing pressure separated by diaphragms. In the last chamber ($10^{-7}$ – $10^{-6}$ mbar), a desolvated beam of charged QDots hits the sample, which is connected to the measurement electronic for continuous monitoring of the tunnel current. Electrostatic deflectors installed in the chamber are used to center the QDot beam on the sample. c) Scanning Electron Microscopy image of one of the 32 pairs of electrodes separated of a distance ~10 nm on which the QDots are projected. d) The tunnel current across these 32 nanogaps is measured simultaneously during the projection of the QDots on the chip circuit ($V_{Drain}$ = 0:1 V, $V_{Gate}$ = 0 V, T=300 K). The trapping of a QDot within a nanogap

*leads to a sharp increase of the tunnel current. e) Sketch of the voltage profile along the device. The applied bias is split equally between the two junctions.*

Electrospray devices have already been employed to depose large biomolecules,[28] polymers,[29,30] metal-organic complex[31] and nanoparticles[28] in several types of experimental studies such as STM,[32] photoemission[33] and optical spectroscopy.[34,35] The electrospray system has been originally developed to produce charged molecular beams for mass spectrometry applications.[36] The main steps of electrospray are : (*i*) The charging of the molecule/nanoparticle solution by the application of a high voltage (≈ 3 kV) on the injector needle. (*ii*) Due to Coulomb repulsions, the formation of a spray of charged droplets at the output of the injector needle, shown in Figure 2b. (*iii*) The evaporation of the solvent as the beam of charged droplets goes through a series of vacuum chambers of decreasing pressure. (*iv*) When all the solvent has evaporated, the naked charged molecules/nanoparticles form a beam into the high vacuum chamber that is directed against the nanogap chip circuit. Additionally, electrostatic deflectors have been installed into the high vacuum chamber to deflect the beam toward the center of the chip circuit.

Before using the QDots, the organic ligands at the surface are exchanged with $S^{2-}$ anions,[37] see methods, to improve the conductance between the QDots and the electrodes, *i.e.* to decrease the height of the tunnel barrier. Just after ligands exchange, the QDot solution is fed into the electrospray, Figure 2a. The QDots are projected until a sharp increase of the tunnel current is observed, Figure 2d, at which point the projection is immediately stopped. As discussed in Ref. 27, this sharp increase of the tunnel current is the signature of the trapping of a single QDot within the nanogap. The sample chamber is then dismounted and transferred into a glovebox from which the chip circuit is taken out. In this glove-box, a cryofree cryostat, with $T_{base}$ ≈ 5 K, is installed, on which the chip circuit is mounted. Thus, the sample is never exposed to air during junction fabrication.

Out of the several chips of 32 nanogaps, two samples have been selected because of their large conductance. The differential conductance is measured with a standard lock-in configuration, where an AC voltage, $\Delta V_{AC}$, is added to the continuous DC voltage at the drain electrode. The current output at the source electrode is fed into a current-voltage converter with a gain of $10^9$ V/A, the output of which is measured with a lock-in and standard DC voltmeter to obtain the AC, $\Delta I_{AC}$, and DC contributions to the current. The differential tunnel conductance is obtained directly from the ratio $g_T = dI/dV = \Delta I_{AC}/\Delta V_{AC}$. The gate voltage is applied on the doped silicon substrate. Figure 3a-b shows a color map of the differential tunnel conductance $g_T=dI/dV$ as function of drain and gate voltages for sample A, measured at *T=80 K*. The data for a second sample, sample B, are shown in Figure S2. The small quantitative difference can be attributed to the nanoparticle size difference and difference of coupling with the nanogap electrodes

These conductance maps present distinct regimes corresponding to different electron filling *N* of the electronic levels, which are sketched Fig. 3e. At the most negative voltage, where *N=0*, the Fermi level is located between the valence and the conduction bands, the differential conductance shows a large gap of amplitude $\Delta V$ ≈ *2 V*. Because the applied voltage is split equally between the two junctions formed by the two electrodes with the QDot, sketched in Figure 2e, *i.e.* the lever arm is $\eta \approx 0.5$, the measured voltage gaps correspond to energy gaps of amplitude $\varepsilon \approx 1$ eV.

Upon increasing the gate voltage, the Fermi energy approaches the electron level *1Se*, this leads to a reduction of the gap amplitude as indicated by the black dash lines. At gate voltage $\approx 0$ V, where the black dash lines crosses, the gap closes. This behavior is the consequence of the *1Se* state crossing the Fermi energy. At this gate voltage, the *1Se* level is simply occupied, *N=1*. In this regime, a Coulomb diamond pattern, Fig. 3d, is theoretically expected but has not been resolved in the experimental data though. Upon increasing the gate voltage, the *1Se* level becomes doubly occupied and, consequently, a gap opens in the excitation spectrum, reaching a value $\Delta V \approx 0.81$ V. As we take into account for the 0.5 lever arm, this leads to a gap of $\approx 0.4$ eV. In this regime *N=2*, the Fermi level is between the two excited electron levels *1Se* and *1Pe*.

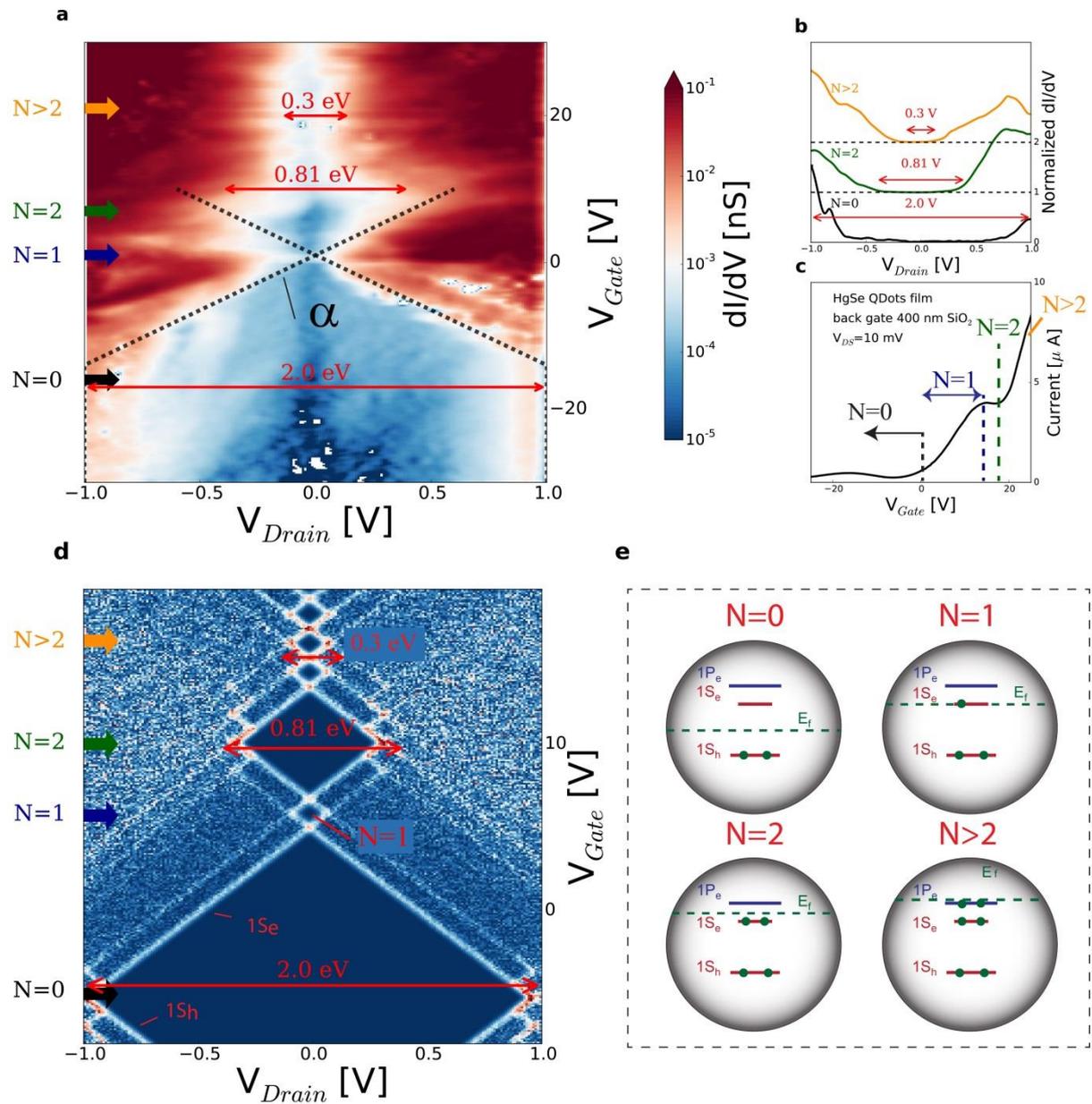

*Figure 3: a) Log scale color map of the differential conductance dI/dV as function of drain and gate voltages for the sample A. The black dash lines are a guide to eyes showing the closing of the gap as the 1Se level crosses the Fermi energy. The four different filing regimes, sketched panel e), are indicated on the left. b) Differential conductance dI/dV as function of drain voltage extracted from the color map a) at the gate voltages corresponding to three distinct QDot filing. c) Current as function of gate voltage for a FET whose channel is made of a thin film of HgSe QDots. On this curve, the different filling regimes can also be identified. d) Simulation of the tunneling spectrum for a QDot. The different regimes are also indicated on the left. e) Sketch of the different occupation levels of the QDot as the gate voltage is increased. The Fermi level is represented by a green dash lines, the electron occupation by green dots.*

From these tunneling gap energies (1 eV and 0.4 eV), one can extract the excitation gap $Eg_0$ in the band structure. This excitation gap is related to the tunneling gap $Eg$ through the relation $Eg = Eg_0 + 2\Sigma$, where $\Sigma \approx 0.078\ eV$ is the Coulomb polarization energy,[38] calculated from the formula $\Sigma = \frac{e^2}{2r}\left(\frac{1}{\kappa_m} - \frac{1}{\kappa_{HgSe}}\right) + 0.47\frac{e^2}{r\cdot\kappa_{HgSe}}\frac{\kappa_{HgSe}-\kappa_m}{\kappa_{HgSe}+\kappa_m}$, with the QDot diameter $2r\approx 10\ nm$, $\kappa_m = 4\pi\varepsilon_m\varepsilon_0$ with $\varepsilon_m = 1.8$, which is the average dielectric coefficient of the media surrounding the QDot and $\kappa_{HgSe} = 4\pi\varepsilon_{HgSe}\varepsilon_0$ where $\varepsilon_{HgSe}= 25.6$ is the static dielectric coefficient of HgSe.**Erreur ! Signet non défini.** From the slope of the excitations lines, highlighted by dashed lines in Figure 3a, one can extract the lever arms $\alpha_{D,S,G}=C_{D,S,G}/C_\Sigma$ for the electrodes Drain, Source, Gate, which provides the corresponding capacitances $C_{D,S,G}$ between the electrodes and the QDot.[39] We find the values $C_D \approx C_S = 1.5\times 10^{-18}\ F$ and $C_G \approx 1\times 10^{-19}\ F$, which are reasonable values for a single QDot capacitively coupled to the electrodes.

For the regimes $N=0$ and $N=2$, ones find excitation gap of $\approx 0.85$ eV and $\approx 0.25$ eV, respectively. These values are remarkably consistent with the interband and intraband gaps values obtained from optical absorption experiment, shown in Figure 1c. This correlation confirms that the two devices are based on a single QDot. Furthermore, because the gap closes near zero gate bias, *i.e.* the regime N=1 is located near zero gate bias, this implies that the mercury chalcogenides QDots are already doped even at near zero gate bias. Moreover it demonstrates that doping exists at the single particle level and is not resulting from a collective effect.

To further confirm our interpretation of the tunneling data, we simulated numerically[40] the expected spectrum for a QDot with states $1S_h$, $1S_e$ and $1P_e$ separated by the energies $1S_h$-$1S_e \approx 0.85$ eV, $1S_e$-$1P_e \approx 0.25$ eV, Coulomb energy $\approx 80$ meV, tunnel resistances $R_1=R_2=10^9\ \Omega$ for the two junctions and using the capacitance values extracted above. The result is shown in Figure 3d. We find that the overall pattern of the experimental spectrum can be reproduced, in particular, the level crossing and the regimes of distinct energy gaps. Note that the different charge states observed in the simulation, leading to the Coulomb diamond patterns at $N=1$ and $N>2$, are not visible in the experimental data. It should also be noticed that the spectrum is symmetric as expected for a symmetric double junction where electron and hole levels are observed both for positive $V_{Drain}>0$ and negative $V_{Drain}<0$. This matter is discussed in the supplementary information.

Interestingly, the intraband gap regime (*N=2*) has also been identified in $I_{Drain}$-$V_{Gate}$ characteristic, shown in Figure 3c, measured on a Field Effect Transistor (FET) whose channel of size 2000x20 µm was made of a thin film of HgSe QDots.[17,19] In this ensemble measurement, a current minimum is observed while the filling of the band reached a value of two carriers per dot. The interpretation for this phenomenon is that the minimum occurs for an average carrier density of 2 per QDot. As the Pauli principle prevents the addition of more carriers, transport has to be thermally activated through the *1P$_e$* state which cost thermal energy and results to this mobility edge. So far this interpretation was only supported by optical measurement of the states filling.[41] Here we demonstrate that the QDot states are indeed involved in transport and that the mobility edge can be observed down to the single particle level.

Finally, at still higher gate voltage, the Fermi level reaches the *1P$_e$* level. Here, many states become accessible (the *1P$_e$* state is 6 times degenerated) that cannot be distinguished and only a narrow gap, ≈ 0.3 V (*i.e.* 0.15 eV in energy with the lever arm $\eta = 0.5$), is observed. The polarization Coulomb gap is at the origin of this narrow gap in the *N>2* regime observed at large positive gate voltage, as well as the residual gap observed at filling *N=1* where the Fermi level is located on the *1S$_e$* state.

Given the importance of photodetection for this material,[16,17] we finally decided to probe the phototransport of a single nanoparticle. We build an experimental setup which enables the simultaneous measurements of the tunnel conductance and the photoresponse. The challenge comes from the fact that in infrared material the photocurrent is only a small modulation of the dark current due to the high thermally activated carrier density. Thus we choose to modulate the drain voltage signal and the light power illumination at two different frequencies and extract the two signals thanks to lock-in detection. In practice the drain bias is modulated at the frequency $\omega_0$ = 17 Hz. The sample illumination is ensured by a LED (*λ*=660 nm - P≈80 mW.cm$^{-2}$) which input is electrically modulated at the frequency $\omega_1$ = 52 Hz. A scheme of the setup is given in Figure 4a. The tunnel current is proportional to the drain bias, $I_{Tunnel}$ α $g_T$ cos ($\omega_0 t$), while the photo-current is proportional to $I_{Photo}$ α $g_P$•cos ($\omega_0 t$) •cos ($\omega_1 t$) α $g_P$ cos ($\omega_0 + \omega_1$)t. Thus while the tunnel current is measured at the frequency $\omega_0$ = 17 Hz, the photo-current is measured at the frequency $\omega_0$+ $\omega_1$=69 Hz. One could observe that the fourth harmonic of the tunnel signal (4x17Hz=68 Hz) has only 1 Hz difference with the photo-current signal. However, using for the lock-in a time constant of about 5 s and 24 dB low-pass filter leads to a bandwidth of 0.023 Hz, which is much smaller than this 1 Hz difference, thus there is no doubt on the origin of the measured photo-current signal.

As expected, the tunnel current (*i.e.* signal at 17Hz) presents only a very limited modulation under illumination. Figure 4d shows that the tunnel conductance without illumination (dash orange line) is almost identical to the tunnel conductance signal under illumination (continuous orange line). On the other hand the amplitude of the photo-current signal, measured at 69 Hz, is enhanced by a factor of two with respect to the signal in the dark, see dark curve in Figure 4d. We actually attribute the background (*i.e.* dark signal) signal at 69 Hz to anharmonic effect. Because of the non-linear IV curve of the system, frequency mixing occurs and leads to a contribution at 69 Hz even in the dark. Nevertheless this parasitic signal is clearly below the contribution in presence of light. The color map, Figure 4c, shows that the evolution of the photo-conductance signal clearly follows the evolution of the tunnel conductance signal, see Figure 4b.

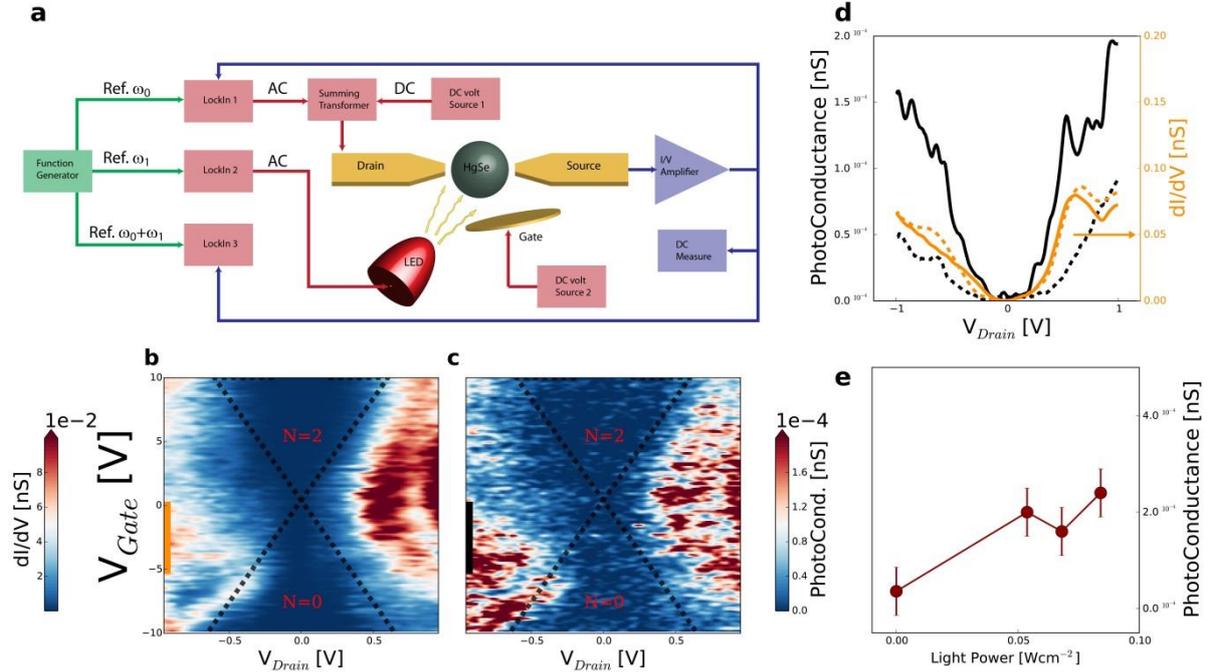

*Figure 4 a) Sketch of the electronic circuit used to extract the photo-conductance signal from the dark conductance. The drain voltage is modulated at the frequency $\omega_0$, the LED is modulated at the frequency $\omega_1$, the conductance signal is extracted at the frequency $\omega_0$, the photo-conductance signal is extracted at the frequency $\omega_0+\omega_1$. The differential conductance, panel b), and photo-conductance, panel c), are measured simultaneously with the circuit shown panel a). These maps measured at temperature T ≈ 80 K are taken at drain-gate voltages corresponding to the 1Se level crossing, shown in Figure 3a. d) Photo-Conductance (black lines) and conductance (orange lines) measured under illumination (continuous lines) and in the dark (dash lines). A large increase in the photo-conductance signal is observed under illumination. The overall pattern of the photo-conductance map is similar to the pattern of the conductance map, indicating the exciton can only be separated at energy where electron tunneling transport is possible. e) Photo-conductance as function of light power.*

Measuring the photoconductance signal as function of light power, Fig. 4e, we found that the photo-conductance is weakly dependent on light intensity in this power range. Thus, assuming that the photocurrent has reached its maximum value, the relation $I_{Photo} = 2e/\tau$ enables to extract the lifetime $\tau \approx 1.6$ µs for the exciton in the QDot, assuming a current $I_{Photo} \approx 0.2$ pA at the drain voltage $V_{Drain} = 1$ V. The lifetime for exciton recombination in mercury chalcogenide QDots is usually below 1 ns.[42,43] A lifetime of ≈1 µs usually suggests that deep traps are involved.

An estimation of the quantum efficiency of the QDot can also be extracted from the relation : $\eta_{ext}= I_{Photo}/eN_{photons}$ where the number of incident photons is given by $N_{photons}=P_{density}$ x $\sigma_{abs}/h\nu$. At 660nm, we estimate the absorption cross section to be equal to 8±2x10$^{-14}$cm$^2$, see supplementary information. This leads to the generation of ≈55 electrons per absorbed photon. Actually such large gains are commonly observed in nanoscale devices[44,45] and results from the fact that the gain is inversely proportional to the transit time (*i.e.* time for the carrier to go from electrode to the other). At 660 nm, the optical pumping leads to interband absorption. Since the material is n-type, the electron is easily transported, while the hole gets trapped for 1 µs according to the estimation of the lifetime. Meanwhile the electron can recirculate 50 times to keep the system neutral. This mechanism called photogating is responsible for the large gain observed.

**Conclusion**

To summarize, we have shown that electrospray deposition of QDots with real time monitoring of the tunnel current provides a reliable way of fabrication single nanoparticle circuits. This enabled a measure of the electron tunneling spectrum consistent with optical and electrical ensemble measurements. Our measurements confirm that the transport occurs through the quantum states of a single QDot and that doping is existing in HgSe self-doped nanocrystal down to the single particle level. Furthermore, on the same junctions, we demonstrate that the photo-conductance of a single colloidal QDot can be measured. Doping atoms or natural surface defects can be spin dependent recombination centers, as shown by the recent observation of spin dependent fluorescence in CdSe/CdS QDots.[46,47] These last works imply that the photo-conductance of QDots should also be spin-dependent, thus providing a possible way toward electrically detected single spin resonance.

**Methods**

**Chemicals**

Se powder (Sigma-Aldrich, 99,99%), mercury acetate (Hg(OAc)$_2$, Sigma-Aldrich), Trioctylphosphine (TOP, Cytek, 90%), dodecanthiol (DDT, Sigma Aldrich), oleic acid (Sigma-Aldrich, 90%), oleylamine (Acros, 80-90%), LiClO$_4$ (Aldrich 98%), Polyethylene glycol ($M_W$=6kg.mol$^{-1}$), N-methylformamide (NMFA, VWR, 98%), Na$_2$S (Aldrich) All chemical are used as received, except oleylamine which is centrifuged before used.

**Nanocrystal synthesis**

The nanocrystals are synthetized according to the procedure given in Ref 17. Briefly, in a three neck flask, 500 mg of mercury acetate are mixed with 10 ml of oleic acid and 25 mL of oleylamine. The flask is degassed at 85 °C for 30 min to form a mercury oleate-oleylamine complex. Meanwhile a 1M solution of Selenium complexed with TOP is prepared by mixing 1.54 g of selenium powder in 20 mL of TOP. The atmosphere of the flask is switched to Ar and the temperature raised to 95 °C. 1.6 mL of the Se solution are quickly injected. The solution turns dark and the reaction is conducted for 4 min. Then 1 mL of dodecanthiol is injected in the flask to quench the reaction and the heating mantle is removed. The flask is further cooled using a fresh air flow. Once the solution is returned to room temperature, the content of the flask is split into two tubes and an equal amount of ethanol is added. The solution is centrifuged (typically at 6000 rpm for three minutes). The clear supernatant is trashed and the pellet is dried under air flow, before being redispersed in toluene. The cleaning procedure is repeated three times. Toluene is used as storage solvent.

**Ligand exchange**

A solution of $S^{2-}$ ligands is prepared by dissolving $Na_2S$ in N-methyl formamide (1% in mass). 1 mL of this solution is introduced in a test tube and the QDot solution is added on the top of it. After strong sonication the QDots get transferred to the polar phase. The clear non polar top phase is removed and trashed. Hexane is added to further clean the polar phase. This step is repeated three times. Finally ethanol is added to precipitate the QDots and the solution is centrifuged for 5 min. The formed pellet is finally redispersed in fresh NMFA.

**Material characterization**

Infrared spectrometry is made on a Bruker vertex 70 FTIR with an ATR system. The solution of nanocrystal is simply dropcasted on the ATR diamond. The spectrum are acquired with a 4 $cm^{-1}$ resolution and averaged 32 times. Scanning electron microscopy is made on a FEI Magellan microscope. Transmission electron microscopy is conducted on a JEOL 2010. The nanocrystal solution is dropcasted on a copper TEM grid. The latter is degassed under secondary vacuum overnight before the imaging to remove remaining organic.

**Microfabrication**

For transport measurements on thin films of QDots, the electrodes are obtained from standard lithography process. On a cleaned doped $Si/SiO_2$ (400 nm oxide) wafer, AZ5214 resist is spin coated. The formed film is soft-baked at 110°C for 90 s. The electrodes pattern is then printed on the resist *via* a 2 s UV illumination through a lithography mask. The electrodes are 2 mm long and spaced by 20 µm. The resist is then inverted during a hard baking step at 125°C on a hot plate for 2 min. Then a 40 s UV flood exposure is performed. Finally we use AZ 726MIF as developer, the wafer is dipped for 32 s in the solution which is quenched by dipping the substrate in distilled water. Once the film is dried, it is introduced in a thermal evaporator where 2 nm of Cr and 30 nm of gold are evaporated.

The nanocrystals, capped with their short ligands are then dropcasted on the electrodes on a hot plate at 100 °C in an air free glove box. While the film is drying, the electrolyte is melted on the same hot plate, and then brushed on the top of the film. A metallic grid is then deposited on the top of the electrolyte to be used as gate electrode. Electrical characterization are conducted in air at room temperature, on a home-made probe station. The latter is connected to two Keithley 2400 sourcemeters used as bias sources and current probes.

Nanogap chips circuits are fabricated by standard e-beam nanofabrication techniques in a clean room. The electrodes Cr (5 nm) /Au (25 nm) are deposited on a p-doped silicon substrate covered with a 300 nm thick silicon oxide layer. One chip of size 8x8 mm contains 32 nanogaps, *i.e.* 32 drain electrodes separated from the common source electrode by a nanogap whose size is in the 7-20 nm range. The electrodes are patterned by e-beam lithography of a PMMA resist on a Vistec EBPG5000+, followed by the thermal evaporation of the electrodes and lift-off. More detail about the chip are given on Figure S1.

**Acknowledgement**

We thanks C. Delerue for useful discussions. We acknowledge the use of clean room facilities from the "Centrale de Proximité Paris Centre". We acknowledge support from ANR grant "QUANTICON" 10-0409-01 and ANR "Nanodose", the Region Ile-de-France in the framework of DIM Nano-K and the Chinese Scholarship Council. This work is also supported by a public grant overseen by the French National Research Agency (ANR) as part of the "Investissements d'Avenir" program (reference: ANR-11-IDEX-0004-02, labex MATISSE)

**Supporting Informations**

Additional data concerning nanogap chip, the tunnel spectrum of sample B and the optical cross section determination are given in supplementary. This material is available free of charge *via* the Internet at http://pubs.acs.org/.

**TOC figure**

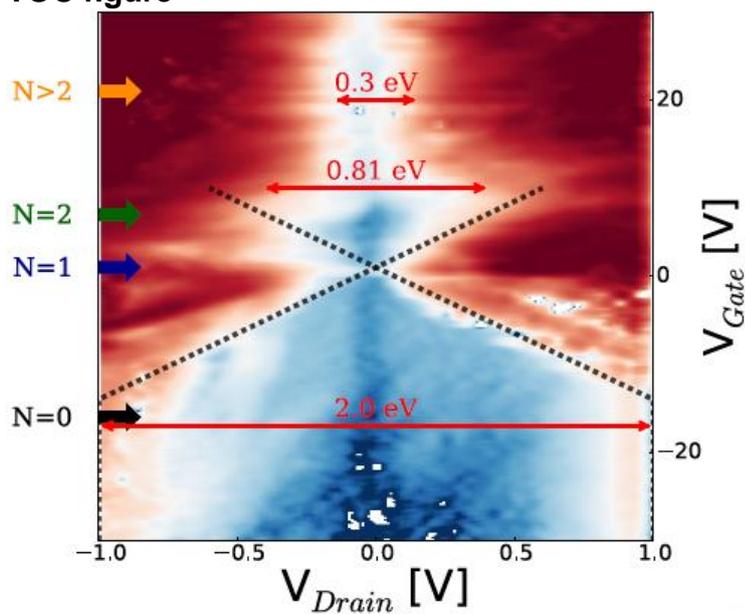

Tunnel spectrum of a single HgSe quantum Dot